\documentclass[letterpaper,10pt]{article}

\usepackage{amsmath}
\usepackage{amssymb}
\usepackage{psfrag}
\usepackage[dvips]{graphicx}
\usepackage{epsfig}

\newcommand{\dirac}{{\slash \negthinspace \negthinspace \negthinspace \nabla}}
\newcommand{\dd}{\textrm{d}}

\newcommand{\re}{{\mathbb{R}}{\mathrm{e}}}

\author{S.\ Estrada-Jim\'enez,\thanks{sestrada@unach.mx} \,\,\, J.\ R.\ G\'omez-D\'iaz \\
Centro de Estudios en F\'isica y Matem\'aticas B\'asicas y Aplicadas.\\
Universidad Aut\'onoma de Chiapas.\\
Ciudad Universitaria. Carretera Emiliano Zapata Km.\ 8.\\
Tuxtla Guti\'errez, Chiapas, M\'exico. \\
C.\ P.\ 29050 \\
\and
A.\ L\'opez-Ortega\thanks{alopezo@ipn.mx} \\ 
Departamento de F\'{\i}sica. Escuela Superior de F\'{\i}sica y Matem\'aticas. \\
Instituto Polit\'ecnico Nacional. \\
Unidad Profesional Adolfo L\'opez Mateos. Edificio 9. \\
M\'exico, D.\ F., M\'exico. \\
C.\ P.\ 07738 }

\title{Quasinormal modes of a two-dimensional black hole}

\begin{document}

\maketitle

\begin{abstract}
 
For a two-dimensional black hole we determine the quasinormal frequencies of the Klein-Gordon and Dirac fields. In contrast to the well known examples whose spectrum of quasinormal frequencies is discrete, for this black hole we find a continuous spectrum of quasinormal frequencies, but there are unstable quasinormal modes. In the framework of the Hod and Maggiore proposals we also discuss the consequences of these results on the form of the entropy spectrum for the two-dimensional black hole.  

\vspace{.3cm}



\end{abstract}

\section{Introduction}
\label{s: Introduction}

The behavior of test fields in curved spacetimes  is thoroughly studied. Among the more analyzed characteristics we find the damped oscillations of a test field propagating in a black hole usually known as quasinormal modes (QNM) \cite{Kokkotas:1999bd}--\cite{Konoplya:2011qq}. These modes have a set of well defined complex frequencies commonly called quasinormal frequencies (QNF) \cite{Kokkotas:1999bd}--\cite{Konoplya:2011qq}.

Recently the QNF of several lower dimensional spacetimes have been calculated \cite{Birmingham:2001pj}--\cite{LopezOrtega:2011sc}. For many of these backgrounds the QNF are exactly computed. This fact allows us to study in some detail some recent proposals involving the QNM \cite{LopezOrtega:2011np}.

A two-dimensional analogue of the Einstein equations for vanishing cosmological constant is given by \cite{Brown:1986nm}--\cite{Landsberg:1993eh}
\begin{equation} \label{e: equation of motion gravity}
 R = 8 \pi T,
\end{equation}
where $R$ is the Ricci scalar and $T$ is the trace of the energy-momentum tensor. 

Several solutions to the equations of motion (\ref{e: equation of motion gravity}) are known \cite{Brown:1986nm}--\cite{Landsberg:1993eh}. In particular for a stationary point mass located at $r=0$ with
\begin{equation}
 T = \frac{M}{2 \pi} \delta (r),
\end{equation} 
it is possible to show that the exterior vacuum solution to the equations of motion (\ref{e: equation of motion gravity}) is
\begin{equation}  \label{e: 2D vacumm metric}
 \dd s^2 = (2 M |r| -C) \dd t^2 - (2 M |r| -C)^{-1} \dd r^2,
\end{equation} 
where $C$ is a constant. Depending on the magnitudes and signs of the quantities $M$ and $C$, we find that the two-dimensional metric (\ref{e: 2D vacumm metric}) represents a black hole or a naked singularity \cite{Brown:1986nm}--\cite{Landsberg:1993eh}.

To have a reasonable Newtonian limit we choose $M > 0$ in what follows \cite{Mann:1991md} and selecting the appropriate units we take $C=1$. For these values of the quantities $M$ and $C$ the metric (\ref{e: 2D vacumm metric}) describes a two-dimensional black hole with line element
\begin{equation} \label{e: 2D black hole metric}
\dd s^2 = (2 M |r| -1) \dd t^2 - (2 M |r| -1)^{-1} \dd r^2,
\end{equation}
and with horizons located at $r_{\pm}= \pm 1/2 M$. The physical properties of this black hole are studied in detail in Refs.\ \cite{Brown:1986nm}--\cite{Landsberg:1993eh}.

To extend the previous results on the QNM of test fields propagating in lower dimensional black holes, in this paper we calculate the oscillation QNF of the two-dimensional black hole (\ref{e: 2D black hole metric}). In the rest of the present work we restrict to the region $r>1/2M$ and therefore we analyze the propagation of the field outside the horizon at $r_+ =+ 1/2 M$. We expect to find similar results in the interval $r<-1/2M$. 

We organize this paper as follows. For the two-dimensional black hole (\ref{e: 2D black hole metric}), in Sect.\ \ref{s: Klein-Gordon field} we determine the QNF of the Klein-Gordon field and in Sect.\ \ref{s: Dirac field} we calculate the QNF of the Dirac field. Finally in Sect.\ \ref{s: Discussion} we use these results to determine the entropy spectrum of the black hole (\ref{e: 2D black hole metric}) and discuss our main results.

\section{QNF of the Klein-Gordon field}
\label{s: Klein-Gordon field}

Since in two-dimensional spacetimes the QNM are well defined only for massive fields \cite{LopezOrtega:2011sc}, in what follows we consider fields with mass different from zero. Denoting with $\square$ to the d'Alembertian and with $m$ to the mass of the field, we write the Klein-Gordon equation as
\begin{equation} \label{e: Klein-Gordon equation}
\left(\square + m^{2}\right) \Phi = 0 .
\end{equation} 
In the two-dimensional black hole (\ref{e: 2D black hole metric}) for $r > 1/2 M$ the Klein-Gordon equation simplifies to the radial differential equation 
\begin{equation} \label{e: radial Klein-Gordon}
\frac{\dd }{\dd r} \left( f \frac{\dd R}{\dd r} \right) + \frac{\omega^2 R}{f} - m^2 R = 0, 
\end{equation} 
when we take
\begin{equation} \label{e: Klein Gordon ansatz}
 \Phi = \textrm{e}^{- i \omega t} R(r),
\end{equation} 
and define 
\begin{equation} \label{e: function f}
 f(r)= 2 M r - 1.
\end{equation} 

Using the tortoise coordinate 
\begin{equation} \label{e: tortoise coordinate}
 r_* = \int \frac{\dd r}{2 M r - 1} = \frac{1}{2M} \ln (2 M r - 1),
\end{equation} 
we transform the radial equation (\ref{e: radial Klein-Gordon}) into the Schr\"odinger type equation \cite{LopezOrtega:2011sc}, \cite{Mann:1991md}
\begin{equation} \label{e: Schrodinger type Klein Gordon}
 \frac{\dd^2 R }{\dd r_*^2} + \omega^2 R = m^2 \textrm{e}^{2 M r_*} R = V R,
\end{equation} 
with the effective potential $V = m^2 \textrm{e}^{2 M r_*}$. Notice that for $r \in (1/2M,\infty) $ we find that $r_* \in (-\infty, +\infty)$. 

Near the horizon and in the asymptotic region the effective potential $V$ behaves as
\begin{equation}
 \lim_{r \to r_+} V = \lim_{r_* \to - \infty} V = 0, \qquad \qquad \lim_{r \to + \infty} V = \lim_{r_* \to + \infty} V \to + \infty .
\end{equation} 
Based on this behavior of the effective potential, for a test field we define its QNM as the solutions to its equations of motion that satisfy the boundary conditions \cite{Kokkotas:1999bd}--\cite{Konoplya:2011qq}
\begin{enumerate}
 \item[i)] Near the horizon the field behaves as $\textrm{exp}{(-i\omega(t + r_*))}$.
\item[ii)] The field goes to zero as $r \to + \infty$.  
\end{enumerate}
 
To solve the Schr\"odinger type equation (\ref{e: Schrodinger type Klein Gordon}) we follow to Mann, et al.\ \cite{Mann:1991md} and make the change of variable
\begin{equation} \label{e: z variable definition}
 z = \frac{m}{M} \textrm{e}^{M r_*},
\end{equation} 
to find that Eq.\ (\ref{e: Schrodinger type Klein Gordon}) becomes \cite{Mann:1991md}
\begin{equation}
 z^2 \frac{\dd^2 R}{\dd z^2} + z \frac{\dd R}{\dd z} - \left[ \left( \frac{i \omega}{M} \right)^2  + z^2 \right]R = 0 ,
\end{equation} 
whose solutions are the modified Bessel functions $I_\nu (z)$ and $K_\nu (z)$, that is \cite{Mann:1991md}, \cite{Abramowitz-book}, \cite{Guo-book}
\begin{equation} \label{e: solutions radial equation}
 R(z) = C_1 I_\nu (z) + C_2 K_\nu (z),
\end{equation} 
with
\begin{equation}
 \nu = \frac{i \omega}{M},
\end{equation} 
and $C_1$, $C_2$ are constants. Notice that $z \to 0$ as $r_* \to - \infty$ and $z \to + \infty$ as $r_* \to + \infty$.

Since the modified Bessel function $I_\nu (z)$ diverges as $z \to + \infty$ \cite{Abramowitz-book}, \cite{Guo-book}, to satisfy the boundary condition ii) of the QNM we must take $C_1 = 0$ in the formula (\ref{e: solutions radial equation}) and therefore we get that the radial function satisfying the boundary condition at the asymptotic region is
\begin{equation} \label{e: radial QNM Klein-Gordon}
 R(z) =  C_2 K_\nu (z).
\end{equation} 
Assuming that the frequencies $\omega$ are complex \cite{Kokkotas:1999bd}--\cite{Konoplya:2011qq}, that is, $\omega = \omega_R + i \omega_I$, where $\omega_R$ ($\omega_I$) denotes the real (imaginary) part of the complex frequency, we find that the parameter $\nu$ takes the form
\begin{equation} \label{e: nu Klein-Gordon}
 \nu = \frac{-\omega_I + i \omega_R}{M} .
\end{equation} 

In what follows we take $\omega_R > 0$. For $\omega_I < 0 $, from the expression (\ref{e: nu Klein-Gordon}) we find that $\re (\nu) > 0$ and hence we use that for small $z$ the modified Bessel function $K_\nu (z)$ satisfies (Eq.\ (9.6.9) of Ref.\ \cite{Abramowitz-book})
\begin{equation} \label{e: property modified Bessel near 0}
 K_\nu (z) \approx \frac{1}{2} \Gamma(\nu) \left(\frac{1}{2} z \right)^{-\nu} \quad \textrm{as} \quad z \to 0 \quad \textrm{and} \quad \re (\nu) > 0 .
\end{equation} 
Thus for $\omega_I < 0$  and near the horizon ($z=0$) the radial function (\ref{e: radial QNM Klein-Gordon}) behaves as
\begin{equation}
 R \approx C_2 \frac{1}{2} \Gamma(\nu) \left(\frac{1}{2} z \right)^{-\nu} 
    \approx \textrm{e}^{-i \omega r_*},
\end{equation} 
and since the time dependence of the Klein-Gordon field is of the form $\textrm{exp}(-i \omega t)$,  for all $\omega_I < 0$ the radial function $R$ represents an ingoing wave near the horizon. Thus when $\omega_I < 0$ the radial function (\ref{e: radial QNM Klein-Gordon}) satisfies the boundary condition of the QNM near the horizon and hence the complex frequencies with $\omega_R > 0$ and $\omega_I < 0$ are QNF of the two-dimensional black hole (\ref{e: 2D black hole metric}). 

For $\omega_I > 0$, from the expression (\ref{e: nu Klein-Gordon}) we obtain that $\re (\nu) < 0$, but we use that the modified Bessel function satisfies \cite{Guo-book} 
\begin{equation} \label{e: modified Bessel minus plus}
 K_\nu (z) = K_{-\nu} (z) ,
\end{equation}  
to write the radial function (\ref{e: radial QNM Klein-Gordon}) as
\begin{equation} \label{e: radial function Klein-Gordon wI positive}
 R(z) =  C_2 K_{-\nu} (z),
\end{equation} 
with $\re (-\nu) > 0$. From the property (\ref{e: property modified Bessel near 0}) of the modified Bessel function we find that for $\omega_I > 0$ and near the horizon the radial function (\ref{e: radial function Klein-Gordon wI positive}) simplifies to
\begin{equation}
 R \approx C_2 \frac{1}{2} \Gamma(-\nu) \left(\frac{1}{2} z \right)^{\nu} 
    \approx \textrm{e}^{i \omega r_*},
\end{equation} 
thus, near the horizon the radial function (\ref{e: radial function Klein-Gordon wI positive}) represents outgoing waves when $\omega_I > 0$. Therefore for $\omega_I > 0$ the radial function $R$ does not satisfy the boundary condition of the QNM. 

From these results we assert that in the two-dimensional black hole (\ref{e: 2D black hole metric}) it is possible to meet the boundary conditions of the QNM when the imaginary part of the frequency satisfies $\omega_I < 0$ (and $\omega_R > 0$). Therefore we get a continuum of QNF and since the time dependence is of the form $\textrm{exp}(-i \omega t)$ (see the formula (\ref{e: Klein Gordon ansatz})) these QNM are stable. It is convenient to remark that we do not know other spacetime with a continuum of QNF for the test fields \cite{Kokkotas:1999bd}--\cite{Konoplya:2011qq}.

\section{QNF of the Dirac field}
\label{s: Dirac field}

In this section we calculate the QNF of the Dirac field propagating in the two-dimensional black hole (\ref{e: 2D black hole metric}). In what follows we denote with $\gamma^\mu$ to the Dirac matrices, with $\nabla_\mu$  to the covariant derivative, with $\dirac = \gamma^\mu \nabla_\mu$ to the Dirac operator, with $m$ to the mass of the fermion field, and with $\Psi$ to the two-dimensional spinor
 \begin{equation} \label{e: two spinor} 
 \Psi = \left(\begin{array}{c}
\Psi_{1}\\
\Psi_{2}  \end{array} \right) .
\end{equation} 
Hence the Dirac equation is given by
\begin{equation} \label{e: Dirac equation}
 i \dirac \Psi = m \Psi .
\end{equation}

To simplify the Dirac equation in the two-dimensional black hole (\ref{e: 2D black hole metric}) for $r > 1 / 2 M$ we follow the procedure of Ref.\ \cite{LopezOrtega:2011sc} and hence we define the line element $\dd \tilde{s}^2$ by
\begin{equation} \label{e: definition element of line}
 \dd s^{2} = f\dd t^2 -\frac{ \dd r^{2} }{ f }=  f \left( \dd t^{2}-\frac{ \dd r^{2} }{ f^2} \right) = f(\dd t^2 - \dd r_*^2 ) = f \dd \tilde{s}^2 ,
\end{equation} 
that is,
\begin{equation} \label{e: line element tilde}
 \dd \tilde{s}^2 = \dd t^2 - \dd r_*^2.
\end{equation} 
Thus the line elements $\dd s^{2}$ and $\dd \tilde{s}^2 $ are related by a conformal transformation, and therefore the components of their metrics satisfy
\begin{equation}
 g_{\mu \nu} = f \tilde{g}_{\mu \nu} .
\end{equation} 
We point out that the definition of the function $f$ appears in the formula (\ref{e: function f}) and that of the tortoise coordinate $r_*$ in the expression (\ref{e: tortoise coordinate}). Notice that in the coordinates $(t,r_*)$  the line element $\dd \tilde{s}^2 $ is flat. 

We recall that if we transform the spinor $\Psi$, the Dirac operator $\dirac$, and the mass of the field $m$ in the form \cite{Gibbons:1993hg}, \cite{Das:1996we}, \cite{LopezOrtega:2009qc}
\begin{align} \label{eq: transformations of spinors}
f^{1/4} \Psi=\tilde{\Psi} ,     \qquad \quad
f^{3/4} \dirac \Psi = \tilde{\dirac} \tilde{\Psi},   \qquad \quad
f^{1/2} m = \tilde{m}, 
\end{align} 
then $\tilde{\Psi}$, $\tilde{\dirac} \tilde{\Psi}$, and $\tilde{m}$ satisfy the Dirac equation in the two-dimensional spacetime with line element $\dd \tilde{s}^2 $. 

Using the chiral representation of the gamma matrices \cite{LopezOrtega:2011sc}
\begin{align} \label{eq: chiral gamma matrices}
 \gamma^{0} = \gamma^{t} = \left( \begin{array}{cc}
0 & 1 \\
1 & 0 \end{array} \right),  \qquad \quad
 \gamma^{1}= \gamma^{r_*} =\left( \begin{array}{cc} 
0 &-1 \\ 
1 & 0 \end{array} \right),
\end{align} 
that fulfill
\begin{equation}
 \gamma^{\mu}  \gamma^{\nu} +  \gamma^{\nu}  \gamma^{\mu} = 2 \eta^{\mu \nu} I, 
\end{equation} 
where $\mu, \nu = t, r_*$ and $\eta^{\mu \nu} = \textrm{diag} (1,-1) $, we find that in the two-dimensional spacetime with metric (\ref{e: line element tilde}) the Dirac equation takes the form
 \begin{align} \label{eq: Dirac equation chiral}
\partial_{t} \tilde{\Psi}_{2} - \partial_{r_*} \tilde{\Psi}_{2}= -i m \sqrt{f} \,\,\, \tilde{\Psi}_{1}, \nonumber \\
\partial_{t} \tilde{\Psi}_{1}  + \partial_{r_*} \tilde{\Psi}_{1}= -i m \sqrt{f} \,\,\, \tilde{\Psi}_{2} ,
\end{align} 
with $\tilde{\Psi}_{1}$ and $\tilde{\Psi}_{2}$ denoting the components of the two-dimensional spinor $\tilde{\Psi}$.

Proposing that the components of the two-dimensional spinor $\tilde{\Psi}$ take the form
\begin{equation} \label{eq: Dirac separation of variables}
\tilde{\Psi}_{q}(r_*,t) = \hat{R}_{q}(r_*) \,\, e^{-i\omega t},  \qquad \quad q=1,2,
\end{equation}
we get that the system of equations (\ref{eq: Dirac equation chiral}) simplifies to the coupled ordinary differential equations (see also Eq.\ (27) of Ref.\ \cite{LopezOrtega:2011sc})
\begin{align} \label{e: Dirac coupled}
&\frac{\dd \hat{R}_{2} }{\dd r_*} + i\omega \hat{R}_{2} = i \sqrt{f} m \hat{R}_{1},\nonumber\\
&\frac{\dd \hat{R}_{1} }{\dd r_*} - i\omega \hat{R}_{1} =-i \sqrt{f} m \hat{R}_{2} .
\end{align}

Defining 
\begin{equation}
 \hat{R}_{1} = f^{1/4} R_1, \qquad \qquad \hat{R}_{2} = f^{1/4} R_2,
\end{equation} 
we find that the new radial functions $R_1$, $R_2$ satisfy the coupled ordinary differential equations
\begin{align} \label{e: radial Dirac coupled}
 &\frac{\dd R_{2} }{\dd r_*} + \left(\frac{M}{2} + i\omega \right) R_{2} = i m \textrm{e}^{M r_*} R_{1},\nonumber\\
&\frac{\dd R_{1} }{\dd r_*} + \left( \frac{M}{2} - i\omega \right) R_{1} =-i  m  \textrm{e}^{M r_*} R_{2} .
\end{align}
From these equations we obtain that the functions $R_1$, $R_2$ must be solutions of the decoupled differential equations
\begin{align} \label{e: Dirac decoupled}
 &\frac{\dd^2 R_{1} }{\dd r_*^2} + \left( \omega +   i\frac{M}{2} \right)^2 R_{1} = m^2 \textrm{e}^{2 M r_*} R_{1},\nonumber\\
&\frac{\dd^2 R_{2} }{\dd r_*^2} + \left( \omega -   i\frac{M}{2} \right)^2 R_{2} = m^2 \textrm{e}^{2 M r_*} R_{2}.
\end{align}
Notice that these equations are of Schr\"odinger type. Furthermore the previous equations are similar to those obtained by Mann, et al.\ \cite{Mann:1991md}, but they use a different approach. 

Regarding to the radial equation (\ref{e: Schrodinger type Klein Gordon}) for the Klein-Gordon field, to solve the differential equations (\ref{e: Dirac decoupled}) for the Dirac field we can use the already found solutions of Sect.\ \ref{s: Klein-Gordon field} by replacing $\omega$ with $(\omega + i M / 2)$ ($\omega$ with $(\omega - i M / 2)$) in the radial function for $R_1$ ($R_2$).

To calculate the QNF of the Dirac field, from the results of the previous section, we get that the radial functions satisfying the boundary condition ii) of the QNM at infinity are
\begin{equation}
 R_1 (z) = \hat{C}_1 K_{\nu_1}(z), \qquad \qquad R_2 (z) =  \hat{C}_2 K_{\nu_2}(z),
\end{equation} 
where $\hat{C}_1$, $\hat{C}_2$ are constants and the parameters $\nu_1$, $\nu_2$ are equal to
\begin{align} \label{e: nu values Dirac}
 &\nu_1 = \frac{i}{M}\left( \omega + i \frac{M}{2} \right) = -\frac{\omega_I + M/2}{M} + \frac{i \omega_R}{M},\nonumber\\
&\nu_2 = \frac{i}{M}\left( \omega - i \frac{M}{2} \right) = \frac{-\omega_I + M/2}{M} + \frac{i \omega_R}{M}.
\end{align}

First we study the radial function $R_2$. Supposing that the imaginary part of the frequency satisfies $\omega_I > M/2$, from the expressions (\ref{e: nu values Dirac}) we get that $\re (\nu_2) < 0$, but using the property (\ref{e: modified Bessel minus plus}) of the modified Bessel function we obtain that the radial function $R_2$ takes the form \cite{Guo-book}
\begin{equation}
 R_2(z) = \hat{C}_2 K_{\nu_2} (z) = \hat{C}_2 K_{- \nu_2} (z) ,
\end{equation} 
with $\re (- \nu_2) > 0$ and taking into account the property (\ref{e: property modified Bessel near 0}) of the modified Bessel function we find that near the horizon the radial function $R_2$ behaves as
\begin{equation}
 R_2  \approx \textrm{e}^{i \omega r_*} \textrm{e}^{M r_* / 2}.
\end{equation} 
Thus for $\omega_I > M/2$ and near the horizon, the radial function $R_2$ is an outgoing wave and therefore for $\omega_I > M/2$ we can not satisfy the boundary condition of the QNM near the horizon.

For frequencies with $\omega_I < 0$, from the expressions (\ref{e: nu values Dirac}) we find that $\re (\nu_2) > 0$. Using the property (\ref{e: property modified Bessel near 0}) of the modified Bessel function we obtain that near the horizon the radial function $R_2$ behaves as 
\begin{equation} \label{e: radial function 2 imaginary > 0}
 R_2  \approx \textrm{e}^{-i \omega r_*} \textrm{e}^{-M r_* / 2},
\end{equation} 
that is, for $\omega_I < 0$ and near the horizon the function $R_2$ is an ingoing wave. In a similar way to the Klein-Gordon field, for the component $\Psi_2$ we have a continuum of well defined QNF with $\omega_R > 0$ and $\omega_I < 0$. 

Finally, for frequencies with $0 < \omega_I < M/2 $, from the expressions (\ref{e: nu values Dirac}) we get that $\re (\nu_2) > 0$ and near the horizon the radial function $R_2$ behaves as in the formula (\ref{e: radial function 2 imaginary > 0}). Thus for complex frequencies with $0 < \omega_I < M/2 $, the radial function $R_2$ is an ingoing wave near the horizon of the black hole (\ref{e: 2D black hole metric}), therefore we find a continuum of well defined QNF with imaginary parts in the interval $0 < \omega_I < M/2 $. Since we are using a harmonic time dependence of the form $\textrm{exp}(-i \omega t)$ (see the expression (\ref{eq: Dirac separation of variables})), the QNM with frequencies whose imaginary parts are located in the interval $0 < \omega_I < M/2 $ are unstable and the amplitudes of these modes increase with the time. Hence the component $\Psi_2$ of the Dirac field has unstable QNM when it propagates in the two-dimensional black hole (\ref{e: 2D black hole metric}).

Using a similar procedure we obtain that the component $\Psi_1$ has a continuum of well defined QNF for  $\omega_R > 0$  and $\omega_I < -M/2 $. For the frequencies with $ - M/2 < \omega_I < 0 $ the radial function $R_1$ represents a purely outgoing wave near the horizon, and for the frequencies with $\omega_I > 0 $ we also find that the radial function $R_1$ represents a purely outgoing wave near the horizon. Therefore in the two-dimensional black hole (\ref{e: 2D black hole metric}) for the component $\Psi_1$ of the Dirac field we get a continuum of stable QNF with $\omega_R > 0$ and $\omega_I < -M/2 $. 

Thus, as for the Klein-Gordon field, in the two-dimensional black hole (\ref{e: 2D black hole metric}) we find that the Dirac field has continuous spectrum of QNF, but its component $\Psi_2$ has unstable QNM.  

Since in a two-dimensional spacetime, the Dirac equation transforms into a decoupled pair of Schr\"odinger type equations with effective potentials that are SUSY partners (see for example Eq.\ (31) of Ref.\ \cite{LopezOrtega:2011sc}), we expect that the components $\Psi_1$ and $\Psi_2$ must have isospectral QNF \cite{Cooper:1994eh}, but in a similar way to the gravitational perturbations of the Schwarzschild anti-de Sitter black hole \cite{Cardoso:2001bb} they are not isospectral due to the following fact. 

Considering that the component $\Psi_2$ has well defined QNF when the radial function takes the form $R_2(z)$ $= \hat{C}_2 K_{\nu_2} (z)$ and the parameter $\nu_2$ satisfies $\re (\nu_2) > 0$, from Eqs.\ (\ref{e: radial Dirac coupled}) we get that
\begin{equation}
 R_1 = -i \frac{\dd R_2}{\dd z} - \frac{i}{M z} \left( \frac{M}{2} + i \omega \right) R_2 .
\end{equation} 
Taking into account that the modified Bessel function $K_{\nu_2}$ satisfies (see Eq.\ (9.6.28) of Ref.\ \cite{Abramowitz-book})
\begin{equation}
 \frac{\dd K_{\nu_2} (z)}{\dd z} = -\frac{\nu_2}{z}  K_{\nu_2} (z) - K_{\nu_2 - 1}(z),
\end{equation}
we find that if $R_2$ satisfies the boundary condition of the QNM at infinity, then $R_1$ satisfies such boundary condition at infinity. 

From the property (\ref{e: property modified Bessel near 0}) of the modified Bessel function we obtain that in the interval where  $\re (\nu_2) > 0$ but $\re (\nu_2-1) <  0$, if $R_2$ satisfies the boundary condition of the QNM near the horizon, then $R_1$ behaves as
\begin{equation}
 R_1 \approx \tilde{C}_1 \textrm{e}^{- i \omega r_*} \textrm{e}^{-3 M r_* / 2} + \tilde{C}_2 \textrm{e}^{ i \omega r_*} \textrm{e}^{- M r_* / 2}
\end{equation} 
near the horizon ($\tilde{C}_1$ and $\tilde{C}_2$ are constants). Hence, in the interval $\re (\nu_2) > 0$ but $\re (\nu_2-1) <  0$,  when the radial function $R_2$ satisfies the boundary condition of the QNM near the horizon, the radial function $R_1$ has ingoing and outgoing parts near the horizon. Therefore the function $R_1$ does not satisfy the boundary condition of the QNM near the horizon, and the components $\Psi_1$, $\Psi_2$ can not have isospectral QNF.

\section{Discussion}
\label{s: Discussion}

As far as we know \cite{Kokkotas:1999bd}--\cite{Konoplya:2011qq}, the two-dimensional black hole (\ref{e: 2D black hole metric}) is the first example of a spacetime with a continuous spectrum of QNF for test fields, but from our previous results we find that in contrast to the Klein-Gordon field, the Dirac field has unstable QNM since the component $\Psi_2$ has well defined QNF with imaginary parts that satisfy $\omega_I > 0$, and therefore the amplitude of the field increases with the time. 

In a similar way to higher dimensional black holes, the two-dimensional black hole (\ref{e: 2D black hole metric}) has well defined thermodynamic properties \cite{Mann:1990ci}. For this black hole its Hawking temperature and entropy are equal to
\begin{equation} \label{e: temperature entropy}
 T = \frac{\hbar M}{2 \pi}, \qquad \qquad S = \frac{2 \pi}{\hbar} \ln \left(\frac{M}{M_0}\right),
\end{equation} 
where $\hbar = h / 2 \pi$, with $h$ denoting the Planck constant and $M_0$ is an integration constant \cite{Mann:1990ci}.

It is known that the logarithmic dependence of its entropy on the mass (\ref{e: temperature entropy}) produces the tendency of the two-dimensional black hole (\ref{e: 2D black hole metric}) to fractionalize into units of a fundamental mass \cite{Mann:1990ci}. We think that a relevant problem is to determine whether the instability of the QNM found in the previous section is related to this behavior of the black hole (\ref{e: 2D black hole metric}). 

Recent proposals suggest that the asymptotic QNF of a black hole (those with $|\omega_I| \to \infty $) determine the value of its entropy quantum \cite{Hod:1998vk}, \cite{Maggiore:2007nq}. Hod proposes that the real part of the asymptotic QNF determines the change in the mass of the black hole when it emits a quantum \cite{Hod:1998vk}. In accordance with this proposal, when a black hole emits a quantum the change in its mass is equal to
\begin{equation} \label{e: Hod proposal}
 \Delta M = \hbar \omega_R^{(A)},
\end{equation} 
where $\omega_R^{(A)}$ is the real part of the asymptotic QNF \cite{Hod:1998vk}.

From our previous results, for the two-dimensional black hole (\ref{e: 2D black hole metric}) we have asymptotic QNF for which their real parts  $\omega_R^{(A)}$ can take any positive real value, and therefore, according to the proposal by Hod of the formula (\ref{e: Hod proposal}), the change $\Delta M$ in the mass  of the black hole (\ref{e: 2D black hole metric}) takes positive real values. It is convenient to recall that our computation of the QNM for the Klein-Gordon and Dirac fields is based on the assumption of test fields, thus we think that our result is reliable only for small values of $\omega_R^{(A)}$, that is, for small values of $\Delta M$.

From the formulae (\ref{e: temperature entropy}) we get that for the two-dimensional black hole (\ref{e: 2D black hole metric}) the changes in its entropy and mass satisfy
\begin{equation} \label{e: first law}
 \Delta M = T \Delta S ,
\end{equation} 
and therefore, according to the proposal by Hod \cite{Hod:1998vk}, its entropy changes continuously. Thus the entropy of the two-dimensional black hole (\ref{e: 2D black hole metric}) has a continuous spectrum.

Maggiore \cite{Maggiore:2007nq} modifies the proposal by Hod and suggests that the change in the mass of the black hole when it emits a quantum is equal to
\begin{equation} \label{e: Maggiore proposal}
 \Delta M = \hbar \Delta \omega ,
\end{equation} 
where $\Delta \omega$ is the difference of the quantities $\sqrt{\omega_R^2 + \omega_I^2}$ corresponding to two contiguous asymptotic QNF \cite{Maggiore:2007nq}. Following Maggiore, for the two-dimensional black hole (\ref{e: 2D black hole metric}) we take as the contiguous asymptotic QNF to
\begin{equation}
 \omega_1 = (\omega_R + w_R) + i(\omega_I + w_I), \qquad \omega_2 = \omega_R  + i \omega_I ,  
\end{equation} 
where $ w_R$, $ w_I$ are small quantities that satisfy
\begin{equation}
 |w_R|, |w_I| \ll |\omega_R| \ll |\omega_I|.
\end{equation} 

Hence we find that for the two-dimensional black hole (\ref{e: 2D black hole metric}) the quantity $\Delta \omega$ simplifies to
\begin{align}
 \Delta \omega &= \left( (\omega_R + w_R)^2 + (\omega_I + w_I)^2 \right)^{1/2} - \left( \omega_R^2 + \omega_I^2 \right)^{1/2} \nonumber \\
& \approx  \left( \omega_R^2 + \omega_I^2 \right)^{1/2} \left( 1 + \frac{\omega_R w_R}{  \omega_R^2 + \omega_I^2 } + \frac{\omega_I w_I}{ \omega_R^2 + \omega_I^2 } \right) -  \left( \omega_R^2 + \omega_I^2 \right)^{1/2} \nonumber \\
& \approx   w_I ,
\end{align} 
as $|\omega_I| \to \infty$.

Since $w_I$ is a continuous small quantity, $\Delta \omega$ varies in a continuous way and therefore from the formulae (\ref{e: first law}) and (\ref{e: Maggiore proposal}) we obtain that the entropy of the black hole (\ref{e: 2D black hole metric}) changes continuously, that is, Maggiore proposal predicts that the entropy of the black hole (\ref{e: 2D black hole metric}) has a continuous spectrum.

Hence the Hod and Maggiore proposals produce that the entropy of the two-dimensional black hole (\ref{e: 2D black hole metric}) has a continuous spectrum, as the entropy spectrum found in Ref.\ \cite{Nomura:2005dn} for the regular black hole. Doubtless it is relevant to compare these results for the entropy spectrum with the results given by other methods.

\section{Acknowledgments}

This work was supported by CONACYT M\'exico, SNI M\'exico, EDI-IPN, COFAA-IPN, and Research Projects SIP-20131340 and SIP-20131541. S.\ Estrada-Jim\'enez and J.\ R.\ G\'omez-D\'iaz acknowledge financial support from research grant PROMEP /103.5/08/3291.

\end{document}